\documentclass[namedreferences]{solarphysics}

\usepackage[hyperref, optionalrh,solaromanenum]{spr-sola-addons} 

\usepackage{graphicx}                    
\usepackage{txfonts}
\usepackage{color}                       

\usepackage{multirow}
\usepackage[rightcaption]{sidecap}

\usepackage[normalem]{ulem}

\hypersetup{
    final=true,
    pageanchor=true,
    colorlinks=true,
    breaklinks=true,
    linkcolor=blue,
    citecolor=blue,
    urlcolor=blue,
    pdfpagemode=UseNone,
    pdftitle={Filigree in the surroundings of polar crown and high-latitude filaments},
    pdfauthor={Diercke et al.},
    pdfsubject={Solar Physics},
    pdfkeywords={Chromosphere, Quiet; Granulation; Magnetic Fields, Photosphere; Prominences, Quiescent}}

\newcommand\phm{\phantom{$-$}}
\newcommand\phn{\phantom{0}}
\newcommand\degr{\ensuremath{^\circ}}

\newcommand\arcsec{\ensuremath{^{\prime\prime}}}
\newcommand\arcsecdot{\mbox{$^{\prime\prime}$}\hspace{-0.15cm}.\,} 

\begin{document}

\begin{article}

\begin{opening}

\title{Filigree in the surroundings of polar crown and high-latitude filaments}

%
\author[addressref={aff1,aff2},email={adiercke@aip.de}]{\inits{A.}\fnm{
Andrea}~\lnm{Diercke}\orcid{0000-0002-9858-0490}}
\author[addressref=aff1, corref]{\inits{C.}\fnm{Christoph}~\lnm{Kuckein}\orcid{0000-0002-3242-1497}}
\author[addressref=aff1, corref]{\inits{M.}\fnm{Meetu}~\lnm{Verma}\orcid{0000-0002-7729-6415}}
\author[addressref=aff1, corref]{\inits{C.}\fnm{Carsten}~\lnm{Denker}\orcid{0000-0003-1054-766X}}

%
\runningauthor{Diercke et al.}
\runningtitle{Filigree in the surroundings of polar crown filaments}

\address[id={aff1}]{Leibniz-Institut f\"ur Astrophysik Potsdam (AIP),
                    An der Sternwarte 16,
                    14482 Potsdam, Germany}

\address[id={aff2}]{Universit\"at Potsdam,
              Institut f\"ur Physik und Astronomie,
              Karl-Liebknecht-Stra\ss{}e 24/25,
              14476 Potsdam, Germany}

\begin{abstract}
High-resolution observations of polar crown and high-latitude filaments are scarce. We present a unique sample of such filaments observed in high-resolution H$\alpha$ narrow-band filtergrams and broad-band images, which were obtained with a new fast camera system at the \textit{Vacuum Tower Telescope} (VTT), Tenerife, Spain. The \textit{Chromospheric Telescope} (ChroTel) provided full-disk context observations in H$\alpha$, \mbox{Ca\,\textsc{ii}}\,K, and \mbox{He\,\textsc{i}}~10830\,\AA. The \textit{Helioseismic and Magnetic Imager} (HMI) and the \textit{Atmospheric Imaging Assembly} (AIA) on board the \textit{Solar Dynamics Observatory} (SDO) provided line-of-sight magnetograms and ultraviolet (UV) 1700\,\AA\ filtergrams, respectively. We study filigree in the vicinity of polar crown and high-latitude filaments and relate their locations to magnetic concentrations at the filaments' footpoints. Bright points are a well studied phenomenon in the photosphere at low latitudes, but they were not yet studied in the quiet network close to the poles.  We examine size, area, and eccentricity of bright points and find that their morphology is very similar to their counterparts at lower latitudes, but their sizes and areas are larger. Bright points at the footpoints of polar crown filaments are preferentially located at stronger magnetic flux concentrations, which are related to bright regions at the border of supergranules as observed in UV filtergrams. Examining the evolution of bright points on three consecutive days reveals that their amount increases while the filament decays, which indicates they impact the equilibrium of the cool plasma contained in filaments.

\end{abstract}

%

\end{opening}
\newpage


\section{Introduction}\label{s:intro}


Filaments are common chromospheric and coronal phenomena, which appear at all latitudes. They form at the borders of opposite-polarity magnetic fields \citep{Babcock1955}, \textit{i.e.}, at the polarity inversion line (PIL). The plasma inside the twisted loop-like structures is relatively cool compared to its  surrounding. A typical filament has an elongated spine with two extreme ends. The ends of the filaments reach down to the photosphere, where they are rooted in concentrations of magnetic flux. Barbs branch from the central spine and build footpoints in the photosphere \citep{Martin1998a, Li2013}. Spine and barbs are built up from individual threads tightly packed parallel to each other. Concerning their location on the Sun, we distinguish three types of filaments \citep{Mackay2010}: quiet-Sun  filaments, intermediate filaments, and active region filaments. The first type appears outside of the activity belt and is rooted in the weak magnetic field of the quiet Sun. The sizes range from very small \citep[\textit{e.g.,}][]{Wang2000, Kontogiannis2020} to very large scales spanning over more than half a solar diameter \citep[\textit{e.g.,}][]{Iazev1988, Kuckein2016, Diercke2018}. Active region filaments develop in the vicinity of active regions and are usually smaller, have shorter lifetimes, and possess a stronger magnetic field than the other types \citep{Mackay2010, Kuckein2012a}, whereas intermediate filaments are filaments which cannot be categorized according to one of the other two groups. Polar crown filaments (PCFs) are a special type of quiet-Sun filaments \citep{Leroy1983}. They are tracers of the global solar dynamo and follow the solar cycle in the ``rush-to-the-pole'' \citep{Xu2018, Diercke2019b}. These filaments indicate areas of weak background fields in polar regions \citep{Li2008, Li2010, Panesar2014}. Typical PCFs form as large structures along the PIL in the filament channel and appear as extended and elongated absorption structures on the solar disk \citep{Gaizauskas2001}. Nonetheless, smaller filaments also appear in filament channels. They are fragments of larger filaments, where only the extreme ends of the filaments are visible and the rest of the spine remains not traceable because of weak density and high latitudes \citep{Schmieder2010, Dudik2012}.

Granulation predominantly covers the solar photosphere. The chromosphere above shows more diverse structures, especially in the dominant absorption line of H$\alpha$. Beside filaments, we find fine structures in form of dark, elongated mottles or fibrils, when longer than a few seconds of arc \citep{Dunn1973, Schmieder2001, Rutten2001, Tsiropoula2012}. Mottles form a network above or around supergranular cells. If they appear radially around  a prominent bright core, the structure is  called rosettes, whereby the bright core is related to the underlying magnetic field \citep{Schmieder2001}. In the wings of H$\alpha$, mottles and fibrils are still visible. Far in the wings, where the typical filamentary structures disappear, photospheric granulation becomes visible, but often with a network of bright points within the intergranular lanes. This pattern was first described by \citet{Dunn1973} and named ``filigree'', where  individual elements were referred to as ``crinkles''. Nowadays, the term bright points is more established \citep{Muller2001}. They are the smallest, resolvable features on the Sun. The majority appears in a more circular shape, but many have an elongated appearance \citep{Dunn1974, Muller2001}. Typical bright points have a width of about $0\arcsecdot3$\,--\,$0\arcsecdot35$ \citep{Dunn1973, Kuckein2019} and the elongated ones have a length of up to $1\arcsecdot0$ \,--\,$2\arcsecdot5$ \citep{Dunn1973, Tarbell1990}, which represents likely a chain of bright points but appears as a single structure because of insufficient spatial resolution.

Immediately after discovering these small-scale structures in the H$\alpha$ line wings, a connection was postulated that they are manifestations of magnetic footpoints \citep{Mehltretter1974, Wilson1981}. Bright points represent single magnetic elements, which are characterized by strong concentrations of magnetic flux, even though there is no one-to-one relation between their brightness and the magnetic flux density \citep{Leenaarts2006}. Each bright point represents an elementary flux tube, which is believed to reach nearly vertically from the photosphere to the corona \citep{Tarbell1990, Muller2001}. They are observed in different lines and are mostly studied in the Fraunhofer G-band with a central wavelength of $\lambda_\mathrm{G} = 4305.5$\,\AA\ \citep{Muller2001}. The connection between bright points in the \mbox{Ca\,\textsc{ii}}\,H\,\&\,K lines and H$\alpha$ filigree was already known since shortly after the first description of filigree and was confirmed in other studies since then \citep{Mehltretter1974, Wilson1981, Leenaarts2006}. Bright points may not spatially coincide exactly at different wavelengths, \textit{e.g.}, G-band and \mbox{Ca\,\textsc{ii}} bright points \citep[see][]{Zhao2009}, but they are physically the same phenomenon in G-band, \mbox{Ca\,\textsc{ii}}\,H\,\&\,K, H$\alpha$, and other molecular bands or spectral lines \citep{Muller2001}. \citet{Kuckein2019} compared the appearance and properties of bright points in several wavelength bands covering the blue continuum at 4505\,\AA, \mbox{Ca\,\textsc{ii}}\,H, \mbox{Ca\,\textsc{i}}~$\lambda$10839\,\AA, \mbox{Na\,\textsc{i}}\,D$_2$, and \mbox{Si\,\textsc{i}}~$\lambda$10827\,\AA. Bright points were detected in all wavelengths, expanding in size with height along the magnetic field lines.

\citet{Dunn1973} found bright points are best observed about $+2$\,\AA\ from the line core of H$\alpha$, while \citet{Leenaarts2006} report that bright points are especially well identified in the blue wing of H$\alpha -0.8$\,\AA. According to the latter authors, bright points are less sharp as in G-band but the H$\alpha$ blue wing has a better contrast compared to the surroundings because of line formation properties of the H$\alpha$ line in the photosphere. In other lines, the bright points were also best seen in the blue wing, \textit{i.e.}, \mbox{Na\,\textsc{i}}\,D$_2$ and \mbox{Si\,\textsc{i}}~$\lambda$10827\,\AA\ \citep{Kuckein2019}.

Only few studies of filaments exist with high-resolution observations from ground-based telescopes \citep[\textit{e.g.},][]{Chae2000, Kuckein2012a}, in particular for quiet-Sun filaments \citep[\textit{e.g.},][]{Engvold2004, Lin2005, Kuckein2016} and polar crown filaments \citep[\textit{e.g.},][]{Lin2003}. To the best of our knowledge, no high-resolution studies attempted to connect photospheric bright points in the quiet-Sun at high-latitudes to processes in the chromosphere related to filaments. Throughout the manuscript, we use the term ``bright point'' when we refer to crinkles, \textit{i.e.}, the single elements of filigree. The goal of the present study is to examine bright points in the vicinity of eight polar crown filaments and study their morphology. Do high-latitude and active-region bright points differ? Does the weaker magnetic field close to the poles result in a different morphology of bright points? Will the interface between mixed-polarity and more unipolar fields have an impact on where PCFs form? Finally, how are bright points related to filaments?

First, we introduce briefly the new synchronized CMOS cameras and the observations of polar crown and high-latitude filaments in the H$\alpha$ narrow-band filtergrams and broad-band images (Section~\ref{s:obs}). We refer to observations in the narrow- and broad-band channels as H$\alpha$ filtergrams and H$\alpha$ images, respectively. In the latter, filigree in the surroundings of high-latitude and polar crown filaments are visible. We examine filaments in H$\alpha$ filtergrams, where they are visible in great detail (Section~\ref{s:morphfilament}) and we describe their evolution with context observations in H$\alpha$ and \mbox{He\,\textsc{i}}~10830\,\AA. The morphological appearance of bright points close to the filament is described in  Section~\ref{s:morphfiligree}. We relate the bright points to concentrations of the magnetic field in the quiet-Sun using line-of-sight (LOS) magnetograms and scrutinize their upper atmospheric response in UV~1700\,\AA\ and \mbox{Ca\,\textsc{ii}}\,K images (Section~\ref{s:magfiligree}). In the Discussion (Section~\ref{s:disc}), we contrast quiet-Sun bright points with those found in the activity belt, which are associated with stronger magnetic field concentrations.


\section{Observations}\label{s:obs}

\begin{figure}
\centerline{
\includegraphics[width=1.0\textwidth]{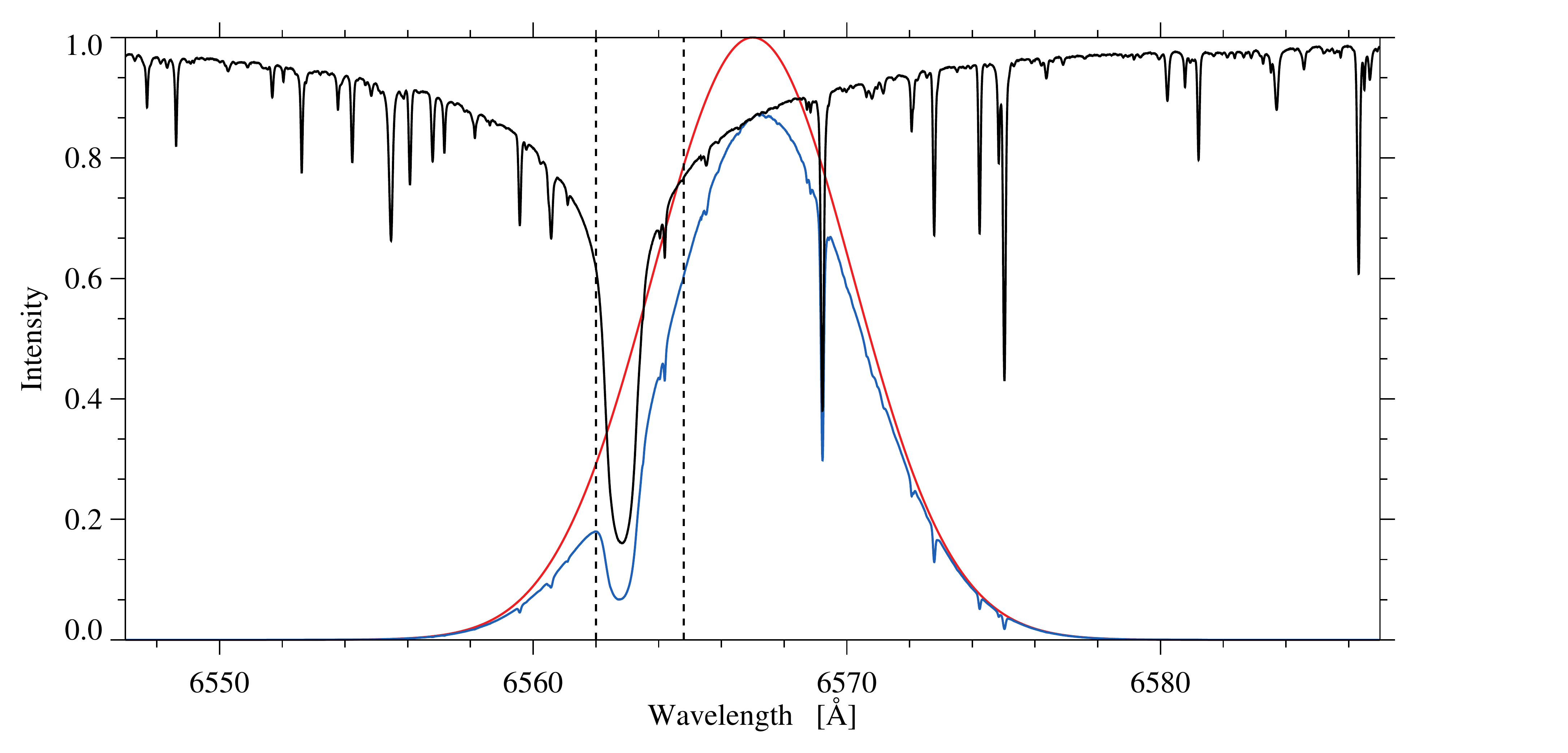}}
\caption{Transmission curve of the H$\alpha$ broad-band filter. Atlas spectrum of the Kitt Peak Fourier Transform Spectrograph (\textit{black}), broad-band filter transmission (\textit{red}), and convolution of both curves (\textit{blue}). The dashed line indicate the positions of H$\alpha -0.8$\,\AA\ and H$\alpha + 2$\,\AA.}
\label{fig:ha_trans}
\end{figure}

Two new synchronized high-cadence CMOS M-lite 2M cameras manufactured by LaVision GmbH G\"ottingen (Germany) were tested in September 2018 at the \textit{Vacuum Tower Telescope} \citep[VTT:][]{vonderLuehe1998}. They replaced the old cameras of the GREGOR \textit{Fabry P\'erot Interferometer} \citep[GFPI:][]{ Denker2010, Puschmann2012} at the GREGOR solar telescope \citep{Schmidt2012}. During the tests, a large variety of polar crown and high-latitude filaments were recorded. We used for the observations a broad-band H$\alpha$ filter 6567\,\AA\ with a full-width-at-half-maximum (FWHM) of 7.5\,\AA\ and a transmission of $T_\mathrm{max}=70$\,\%. The transmission curve of the interference filter is plotted in Figure~\ref{fig:ha_trans} at its nominal central wavelength, which is displaced from the H$\alpha$ line core towards the red. The same interference filter served as a prefilter for the narrow-band Lyot filter at (6562.8$\pm$0.3)\,\AA\ for the second camera. The light was distributed between both cameras by a 70/30 beamsplitter cube. The Lyot filter was manufactured by Bernhard Halle Nachf. Berlin-Steglitz \citep{Kuenzel1955} and is temperature controlled. The wavelength can be tuned by changing the temperature and the entrance polarizer. The interference filter was somewhat tilted to avoid retro-reflections from its front side, which shifts the transmission peak towards the blue. Thus, the convolution of the filter curve with the spectrum yields higher line-core intensities than those depicted in Figure~\ref{fig:ha_trans}. The spectrum was taken from the spectral atlas of the Kitt Peak Fourier Transform Spectrograph \citep{Brault1985}. In general, the line wings contribute significantly more to the broad-band intensity than the line core so that filigree can be detected. However, the contrast of bright points will be lower compared to more discriminate filters in the line wings, \textit{e.g.}, H$\alpha - 0.8$\,\AA\ or  H$\alpha + 2$\,\AA\ as suggested by \citet{Dunn1973} and \citet{Leenaarts2006}, respectively.

\begin{table}
\caption{Overview of the date, time, and location of the observed high-latitude and polar crown filaments. The labels `a' and `b' refer to the location of the filaments in Figure~\ref{fig:overview}. The filament type refers to polar crown (P) and high-latitude (H) filaments.
}\label{tbl:overview}
\begin{tabular}{ccccccc}
\hline
\multirow{2}{*}{Date} & \multirow{2}{*}{Time [UT]} & \multicolumn{2}{c}{Coordinates [arcsec]} & \multicolumn{2}{c}{Filament} &  \multirow{2}{*}{Index}   \rule[-2pt]{0pt}{10pt} \\
     &       & $x$ &  $y$  & Label & Type & \\
\hline
21 Sept. 2018  & 08:33:58 -- 08:39:16 & $-$172.0  & \phm547.2    & A & P & a \\
21 Sept. 2018  & 08:44:00 -- 08:45:45 & \phn$-$81.6  & \phm547.2 & B & P & b \\
22 Sept. 2018  & 09:05:36 -- 09:09:45 & \phm614.4 & \phm297.2    & C & H & --\\
24 Sept. 2018  & 08:33:34 -- 08:39:49 & $-$624.0  & $-$364.8     & D & H & a \\
24 Sept. 2018  & 08:40:01 -- 08:41:35 & \phm201.6 & \phm499.2    & E & P & b \\
25 Sept. 2018  & 08:40:47 -- 08:42:33 & \phm384.0 & \phm528.0    & E & P & a \\
25 Sept. 2018  & 08:49:19 -- 08:51:05 & $-$422.4  & $-$374.4     & D & H & b \\
26 Sept. 2018  & 09:37:46 -- 09:39:55 & \phm518.4 & \phm537.6    & E & P & --\\
\hline
\end{tabular}
\end{table}

The M-lite 2M cameras have a detector size of $1920\times1280$\,pixels with a pixel size of $5.86\times5.86$\,$\mu$m$^2$. The quantum efficiency is about 48\% at H$\alpha$. The cameras are able to take fast time-series of images with a frame rate of up to 100\,Hz at 12\,bit.  A programmable timing unit (PTU) was used to synchronize the cameras with an accuracy of about 10\,ns. Furthermore, the images were rapidly stored on Solid-State Drives (SSDs) facilitating image sequences with fast cadence. Details of the camera system are given in \citet[][submitted]{Denker2020}. During the observations, we took sequences of 500\,images with 10 or 20 repetitions, depending on the seeing conditions. The images were taken with an exposure time of $t_\mathrm{exp} = 20$\,ms and an acquisition rate of $f_\mathrm{acq} = 46$\,Hz. The data were recorded with 2$\times$2-pixel binning, which results in an image scale of 0.098\arcsec\,pixel$^{-1}$ and corresponds to about 71\,km on the Sun.

\begin{figure}
\center{
\includegraphics[width=0.98\textwidth]{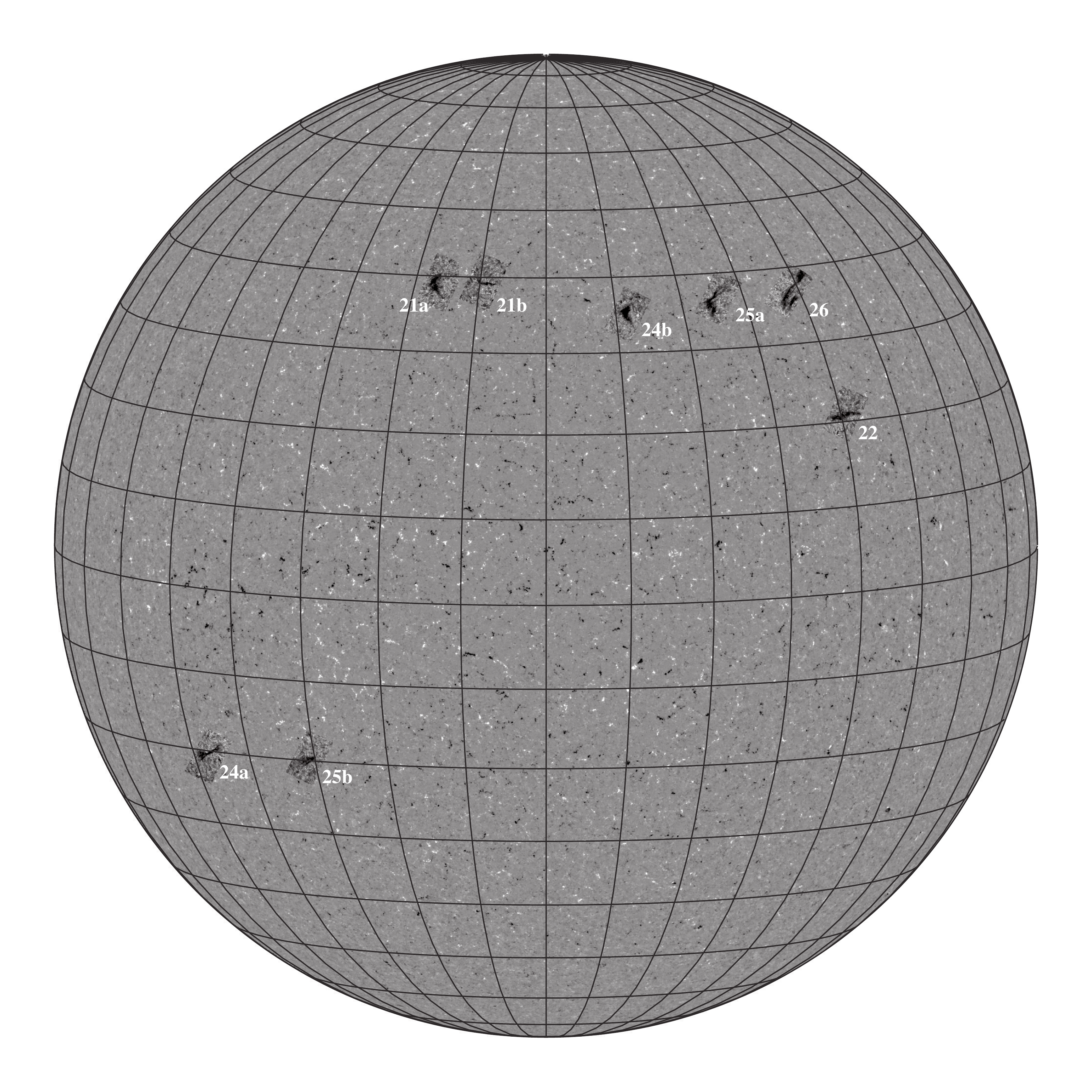}}
\caption{Composite image of a magnetogram and H$\alpha$ filtergrams illustrating the locations of the observed high-latitude and polar crown filaments across the solar disk observed between 21 and 26~September~2018. The corresponding dates and times are collected in Table~\ref{tbl:overview}. Multiple observations on a day are labeled `a' or `b'. The background image is a ``deep'' magnetogram, averaged over 12\,min, observed around 08:34:00~UT on 24~September~2018, which is scaled between $\pm80$\,G. Lines of longitude and latitude are separated by 10\degr.}
\label{fig:overview}
\end{figure}

Using this setup, we observed eight PCFs for five days between 21 and 26~September~2018. The observations were mainly performed in the morning, when seeing conditions are better and more stable. An overview of the location of the filaments is given in Figure~\ref{fig:overview} and the corresponding Table~\ref{tbl:overview}. The dataset consists of five different filaments  (labeled from `A' to `E'), two of which were observed on several days. Most filaments were recorded in the northern hemisphere, and only Filament~D was observed in the southern hemisphere. The locations of the filaments imply that most of the filaments are PCFs, with the exception of Filaments~C and D, which belong to the category of high-latitude filaments.

As a context imager, we use the \textit{Chromospheric Telescope} \citep[ChroTel:][]{Bethge2011} with an aperture of 10\,cm, which provides full-disk H$\alpha$, \mbox{Ca\,\textsc{ii}}\,K, and \mbox{He\,\textsc{i}} 10830\,\AA\ images with a cadence of three minutes. All images went through basic image processing including dark, flat-field, and limb-darkening corrections \citep{Diercke2019b} and were corrected with Zernike polynomials, removing a non-uniform background introduced by the Lyot filters \citep{Shen2018}. We aligned the high-resolution VTT images with the ChroTel H$\alpha$ images and used the ChroTel \mbox{Ca\,\textsc{ii}}\,K images to align the data with the UV 1600\,\AA\ and 1700\,\AA\ filtergrams of the \textit{Atmospheric Imaging Assembly} \citep[AIA:][]{Lemen2012}, which in turn were used to align them with the LOS magnetograms of the \textit{Helioseismic and Magnetic Imager} \cite[HMI:][]{Scherrer2012} on board the \textit{Solar Dynamics Observatory} \citep[SDO:][]{Pesnell2012}. The HMI magnetograms were used to compare the location of bright points, which were derived from the high-resolution broad-band H$\alpha$ images, with their counterparts in the photospheric magnetic field. The SDO data were rescaled and derotated with standard SDO and image processing routines as described in \citet{Diercke2018}.
\newpage


\section{Methods}\label{s:method}


\begin{figure}
\centerline{
\includegraphics[width=1.0\textwidth]{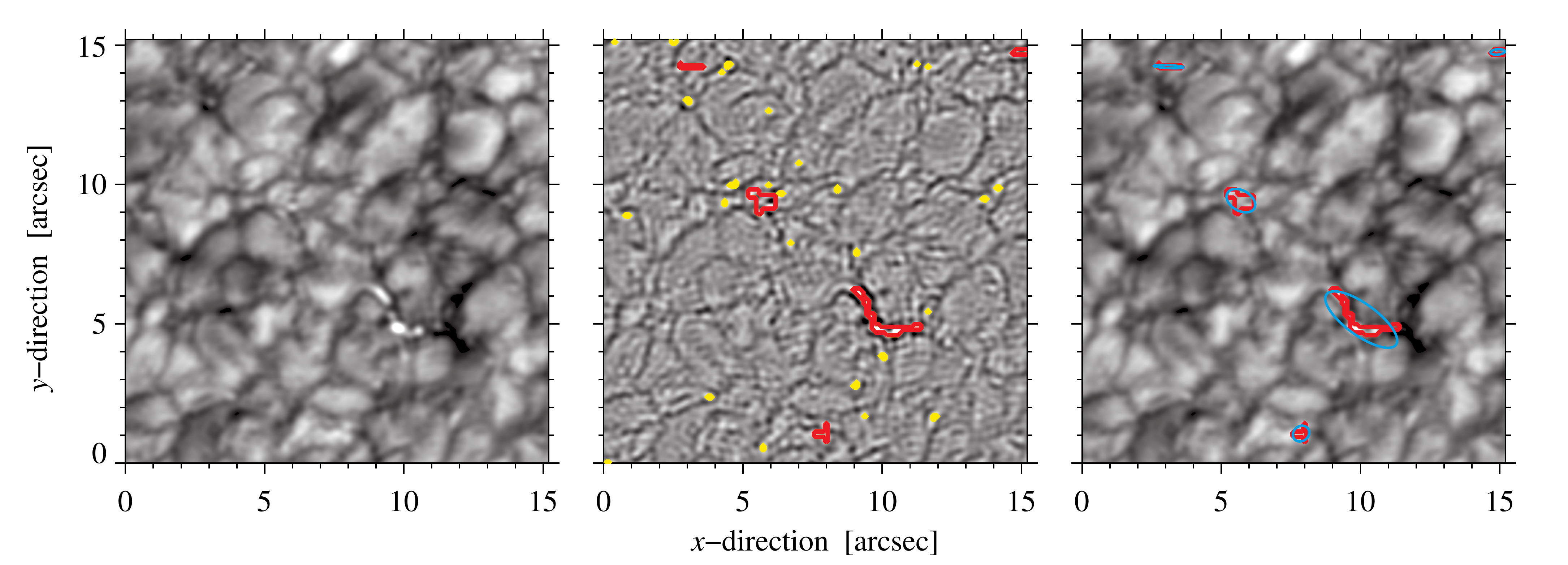}}
\caption{Identification of filigree. Input H$\alpha$ broad-band image  on 21~September~2018 at 08:34:10~UT (Filament~A) for a central region of about $15\arcsec \times 15\arcsec$ (\textit{left}).  H$\alpha$ broad-band image after applying a Laplacian filter with the contours of the pixels above the threshold of three standard deviations in yellow  and the contours of the extracted filigree after applying dilation and erosion  and removing small-scale features in red (\textit{middle}). H$\alpha$ broad-band image with the contours of the extracted filigree in red (\textit{right}). In addition, we display an example for the ellipse fitting, approximating the shape of the bright point (\textit{blue contours}).}
\label{fig:method}
\end{figure}

Aligning data from different instruments is difficult due to large differences in image scale. In consequence, we cannot rely on pixel-to-pixel alignment. We have to take into account the differences in image scale for VTT images (0.098\arcsec\,pixel$^{-1}$), ChroTel images (0.96\arcsec\,pixel$^{-1}$), and SDO magnetograms and UV filtergrams (0.6\arcsec\,pixel$^{-1}$). The low-resolution data were resampled to the same image scale as the VTT data using linear interpolation.

We created ``deep'' magnetograms by averaging 16 magnetograms around the central magnetogram of the time series. Individual HMI magnetograms have an average noise level of about 10\,G with a gradual increase of noise towards the limb \citep{Liu2012}. Considering only photon statistics, the signal-to-noise ratio improves by a factor of four, which is, however, somewhat lower in reality and approaches a limit when the evolution time scale of magnetic features can no longer be neglected. The 12-minute averages are a good compromise between magnetic sensitivity and temporal evolution of small-scale magnetic features in the quiet Sun.

The first step in extracting filigree from broad-band H$\alpha$ images is to normalize the data by dividing the images by the median intensity of the quiet Sun (left panel in Figure~\ref{fig:method}). For identification of bright points, we followed the method presented by \citet{Feng2012} who identified G-band bright points. A Laplacian filter, utilizing the second derivative of the image, helps to obtain candidates for the filigree (middle panel in Figure~\ref{fig:method}). For these candidates, we applied an intensity threshold, which is the mean intensity plus three times the standard deviation of the image (yellow contours in the middle panel of Figure~\ref{fig:method}). We applied morphological dilation to the mask with a cross-shaped  structuring element $s = [[0,\,1,\,0],\,[1,\,1,\,1],\,[0,\,1,\,0]]$. After applying morphological closing to the mask with a 5$\times$5-pixel circular kernel, we removed small-scale features with less than five contiguous pixels. The thus  extracted filigree are shown as red contours in the middle and right panels of Figure~\ref{fig:method}.

A blob finding algorithm  \citep{Fanning2011} identifies contiguous regions  based on the four-adjacency criterion. Several properties follow from the blob analysis, for example, the location of the bright points and their perimeter length and area in pixels. Other properties of bright points can be determined from fitting an ellipse to the structure, for example, the tilt-angle and the semi-major and -minor axis, which gives the eccentricity of the bright points. In the following, we will divide filigree in three classes: bright points in the surroundings of the filaments (Type~I), bright points at the location of the filament (Type~II), and bright points at the footpoints of the filaments (Type III). The footpoints were determined manually by utilizing H$\alpha$ filtergrams of the narrow-band channel and magnetograms. The corresponding locations of the footpoints are marked in subsequent figures by red circles.


\section{Results}\label{s:results}



\subsection{\texorpdfstring{Morphology of the filaments in H$\alpha$}{Morphology of the filaments in H-alpha}}\label{s:morphfilament}

\begin{figure}
\centerline{
\includegraphics[width=0.92\textwidth]{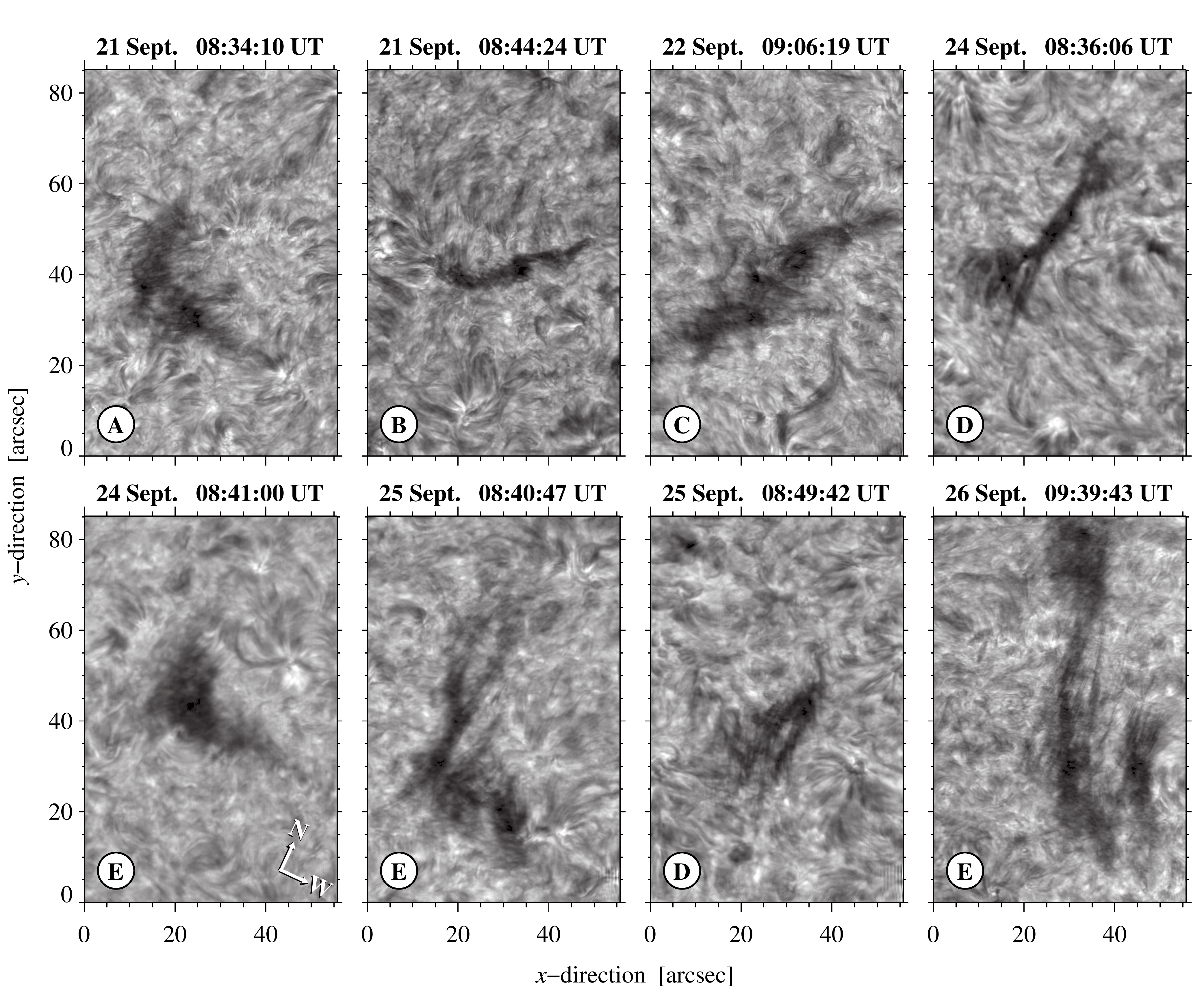}}
\caption{High-resolution H$\alpha$ filtergrams of the polar crown and high-latitude filaments observed with the narrow-band Lyot filter at the VTT. The label of the filaments is noted in a black circle in the lower left corner (see Table~\ref{tbl:overview}). The arrows indicate the direction of solar north and west.}
\label{fig:overview_single}
\end{figure}

All observed filaments are either small-scale polar crown or high-latitude filaments (see Table~\ref{tbl:overview} and Figure~\ref{fig:overview}). We could not record longer time series of PCFs because of unstable seeing conditions. They are small enough to fit into the field-of-view (FOV) of the detectors and are therefore not longer than 80\arcsec\ (Figure~\ref{fig:overview_single}). The morphology of the filaments is clearly seen in high-resolution filtergrams with single threads forming the spines. Some of the filament spines appear as the typical dark elongated structures, \textit{e.g.}, Filament~B in the middle of the FOV. Some filaments are thicker and more diffuse, namely Filaments~A and E. For both filaments, both extreme ends could not be clearly determined, which could be an indication that these were fragments of a larger filament \citep{Schmieder2010}, where only the extreme ends were visible and the rest was not sufficiently dense to be observed in H$\alpha$. Especially strong is this impression for Filament~E on 24 and 25~September. Examining the ChroTel full-disk H$\alpha$ filtergrams (not shown), we see the filament ends for Filament~A and E about 27\arcsec\ to 10\arcsec\ outside the high-resolution FOV. Filament~A covers about 45\,\% of the complete filament, whereas the fraction in the FOV increases for Filament~E from about 15\,\% on 24~September and about 25\,\% on 25~September to about 50\,\% of the complete filament on 26~September. The diffuse impression of these two filaments can be explained by assuming that they are only fragments of a larger filament, whereby for Filament~E on 25 and 26~September single threads become visible along the spine of the filament.

Filament A is only visible on 21~September. During this short period, it is very dynamic. The complete spine between both footpoints becomes visible and we can observe plasma flows from the spine towards the second footpoint in full-disk context observations of the \textit{Kanzelh\"ohe Solar Observatory} \citep[KSO: ][]{Otruba2003, Poetzi2015}. Furthermore, the filament starts to erupt but fails in the process. Afterwards, the filament is only visible as a small-scale structure and dissolves completely. In contrast, Filament E is visible for five days. On the first day, 24~September, the eastern part (Figure~\ref{fig:overview_single}) is very dynamic and changes completely over the course of a day. The small-scale structure develops in an elongated structure, where half of the filament spine becomes visible. Around 14:00\,UT, the filament dissolves again into a small-scale structure. On the second day, 25~September, the spine is more opaque and changes are only small. After 26~September, the filament begins to dissolve. The other filaments B, C, and D are very stable with short lifetimes of one to two days.

In the surroundings of the filaments (Figure~\ref{fig:overview_single}), we see in the filtergrams typical chromospheric fine structures made up of single threads. Some pronounced rosettes, which are formed by radial dark mottles, are anchored in bright cores. Some of these bright cores are related to filigree and strong concentrations of magnetic field. The mottles build a network, which is located around supergranular cells \citep{Schmieder2001}, which is also seen in the H$\alpha$ filtergrams, \textit{i.e.}, Filament~A on the right side of the filament at ($x$, $y$) coordinates (30\arcsec\,--\,55\arcsec, 20\arcsec\,--\,60\arcsec), where the mottles are located around a supergranular cell. Some longer mottles or fibrils are anchored in bright cores and even mini-filaments are seen in the surroundings of the larger polar crown and high-latitude filaments, \textit{i.e.}, Filaments~C and~E on 22 and~24 September, respectively.

\begin{figure}
\centerline{
\includegraphics[width=0.97\textwidth]{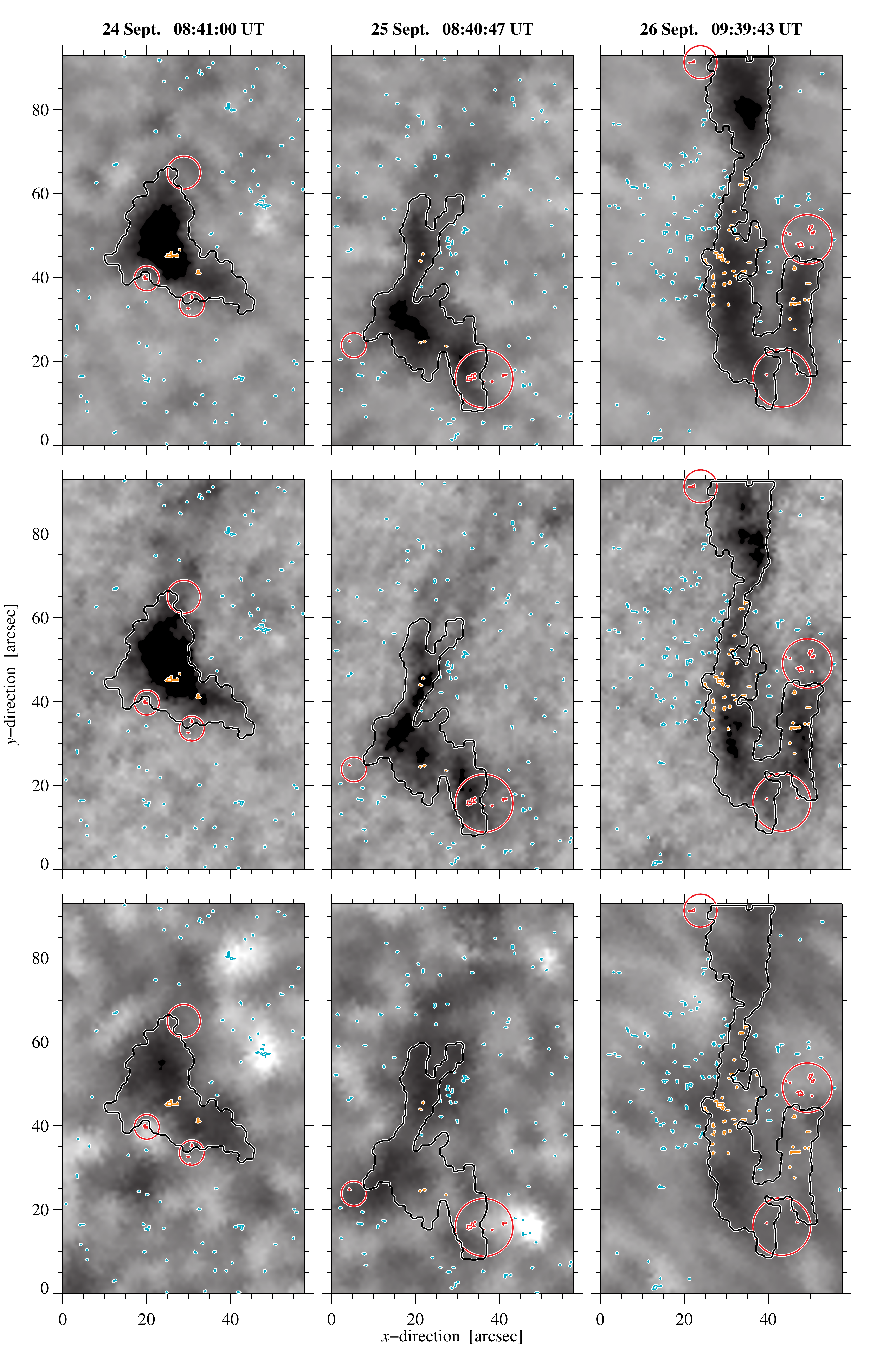}}
\caption{Filament~E observed 24\,--\,26~September~2018 with ChroTel in H$\alpha$ (\textit{top}), in the  red line core of the \mbox{He\,\textsc{i}}~$\lambda$10830\,\AA\ triplet (\textit{middle}), and in \mbox{Ca\,\textsc{ii}\,K}  (\textit{bottom}) with the contours of the filaments derived from the H$\alpha$ filtergrams (\textit{black}) and of filigree in the three categories: Type~I (\textit{cyan}), Type~II (\textit{orange}), and Type~III (\textit{red}). In addition, the footpoints of the filaments are marked with red circles.}
\label{fig:chrotel_compare}
\end{figure}

We compare the appearance of the filaments in the three ChroTel filtergrams using Filament~E as an example (Figure~\ref{fig:chrotel_compare}). In the ChroTel H$\alpha$ filtergrams (upper row in Figure~\ref{fig:chrotel_compare}), Filament~E has a similar morphology as in the high-resolution H$\alpha$ filtergrams but lacks the high contrast fine-structure such as mottles, fibrils, or mini-filaments. The background is uniform and does not provide many details. The shape of the filament is, however, the same when compared to the high-resolution images. On 25~September, we see an extended faint absorption structure in the upper part of the filtergram continuing beyond the contour of the filament between coordinates (20\arcsec\,--\,50\arcsec, 60\arcsec\,--\,90\arcsec).

All filaments are visible in the \mbox{He\,\textsc{i}}~$\lambda$10830\,\AA\ line-core filtergrams (middle row in Figure~\ref{fig:chrotel_compare}). The filaments appear overall very similar as in H$\alpha$ but differences become apparent in a detailed comparison. Strong absorption is encountered in different parts of the filament in \mbox{He\,\textsc{i}} as in the ChroTel H$\alpha$ filtergrams. For example, on 25~September, Filament~E has a stronger absorption in the elongated upper part of the filament in \mbox{He\,\textsc{i}} at coordinates (10\arcsec\,--\,25\arcsec, 35\arcsec\,--\,60\arcsec) but in H$\alpha$ the absorption of the lower, compact part is higher at coordinates (5\arcsec\,--\,35\arcsec, 5\arcsec\,--\,35\arcsec). In addition, on 24 \& 25~September, we see dark, faint absorption structures above Filament~E between coordinates (20\arcsec\,--\,55\arcsec, 60\arcsec\,--\,90\arcsec), which are much fainter in the H$\alpha$ full-disk filtergrams. This strengthens the supposition that these filaments are fragments of larger structures.

The ChroTel full-disk \mbox{Ca\,\textsc{ii}\,K} filtergrams (bottom row in Figure~\ref{fig:chrotel_compare}) show a different chromospheric layer. Nonetheless, some small absorption features are also visible in the \mbox{Ca\,\textsc{ii}\,K} filtergrams at the location of the filaments, for example, Filament~E on 24 and 25~September. This faint \mbox{Ca\,\textsc{ii}}~K absorption in filaments is also reported in, e.g., \citet{Kuckein2016}. On 26~September, the ChroTel \mbox{Ca\,\textsc{ii}}~K filtergram is washed-out, \textit{i.e.}, mediocre seeing conditions prevent a detailed discussion. The network of bright points is associated with supergranular cell boundaries, which are slightly visible around the filaments as bright regions. The enhanced brightness in \mbox{Ca\,\textsc{ii}\,K} filtergrams is closely related to the magnetic field strength \citep{Rutten1991}, and the magnetic field converges at the boundaries of supergranular cells, which are outlined as bright regions in the \mbox{Ca\,\textsc{ii}\,K} filtergrams. In addition, the filigree in the surroundings of the filaments are located close to these regions. This will be discussed in more detail in Section~\ref{s:magfiligree} based on UV~1700\,\AA\ filtergrams.

\begin{figure}
\centerline{
\includegraphics[width=1.0\textwidth]{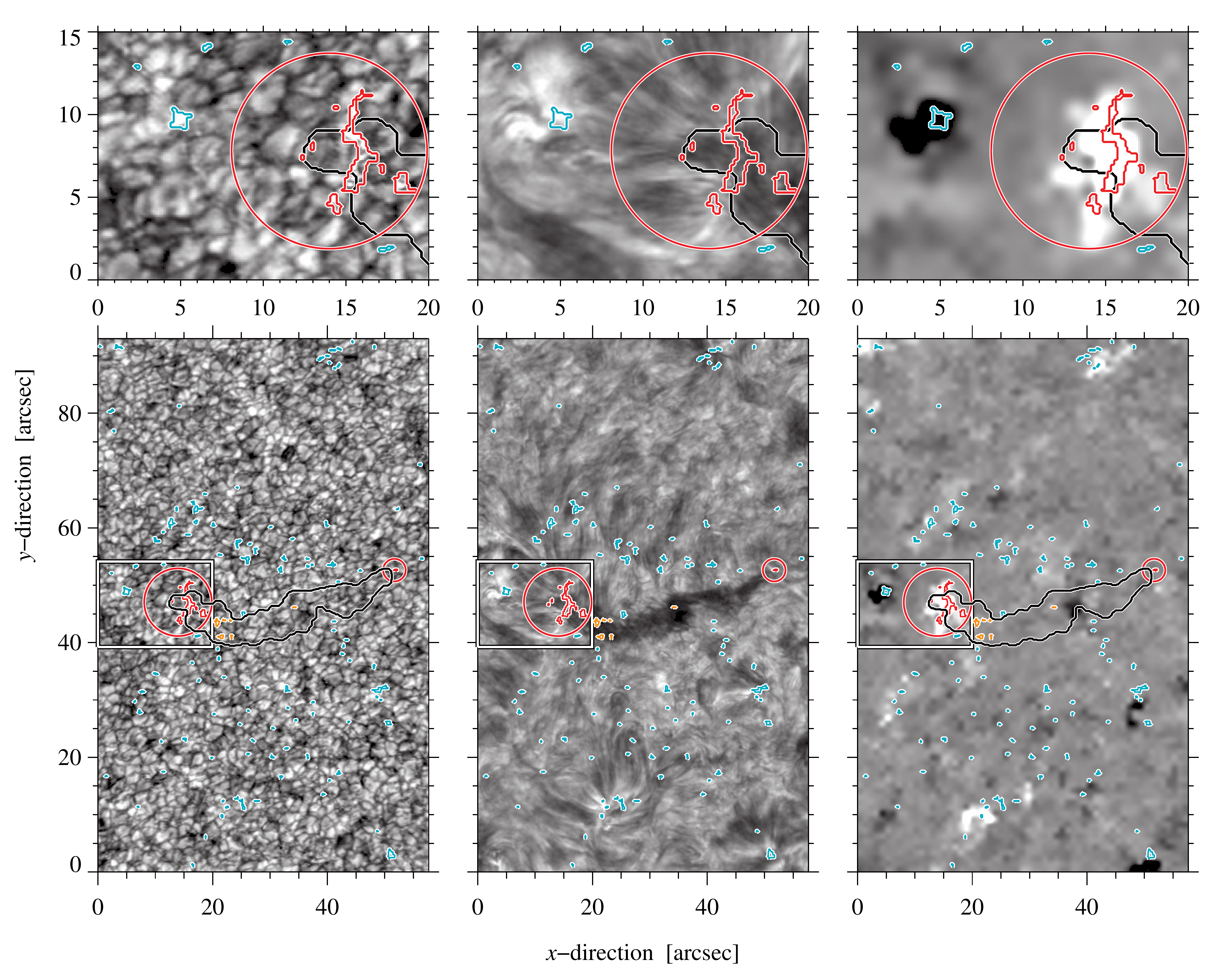}}
\caption{Identified filigree of Type~I (\textit{cyan}), Type~II (\textit{orange}), and Type~III (\textit{red}) for the Filament~B on 21~September~2018 superposed on a high-resolution H$\alpha$ image (\textit{left}) and filtergram (\textit{middle}) and on a corresponding magnetogram (\textit{right}). The deep HMI magnetogram was averaged over 12\,minutes and is displayed between $\pm40$\,G. For clarity, we display the contours of the filament (\textit{black}) and the footpoint regions (\textit{red}). The white rectangle indicates a region-of-interest, which is displayed in the top row at a higher magnification.}
\label{fig:bright}
\end{figure}


\subsection{Morphological description and statistics of filigree} \label{s:morphfiligree}


The broad-band H$\alpha$ images (left panels in Figure~\ref{fig:bright}) show photospheric granulation on 22~September above which Filament~B is suspended. The magnified left footpoint region establishes the location of Type~I and ~II bright points in the high-resolution image of granulation. The granulation exhibits an irregular brightness distribution, \textit{i.e.}, some very bright granules and some darker areas with the size of a granule. Some of these dark areas are likely remnants of strong absorption structures leaving imprints in the broad-band H$\alpha$ images. The bright points are located at the border of the granules, \textit{i.e.}, along the intergranular lanes. Inspecting the H$\alpha$ filtergram (middle panel in Figure~\ref{fig:bright}), we recognize that bright points are located near the footpoints of the filaments (red circles) but also coincide with bright cores of rosettes at coordinates (25\arcsec, 15\arcsec). Fibrils extend between the footpoint of the filament and the bright point to its left. Mapping bright point positions onto magnetograms (upper right panel in Figure~\ref{fig:bright}) reveals that both bright points are rooted at opposite polarities, crossing the polarity inversion line. The larger FOV depict that many, but  not all, bright points are associated with strong magnetic field concentrations. Magnetic field observations with higher spatial resolution may even yield a closer association. Moreover, some bright points are very close to strong magnetic fields but do not exactly match their positions. This may be due to alignment and projection errors, and differences in the three instruments. However, this behavior was also noted in other studies, \textit{e.g.}, for G-band and \mbox{Ca\,\textsc{ii}} bright points \citep{Zhao2009}. The magnetograms will be analyzed in more detail in Section~\ref{s:magfiligree}.

In this example, we already encounter different shapes and sizes of bright points. Visually, the large majority is small and has a more roundish form but some are larger and either roundish or elongated. Some bright points are located in clusters, as seen for Filament~B (right panel in Figure~\ref{fig:bright}) at the upper footpoint, where no single bright point can be distinguished. In the following, we will determine area distribution, perimeter length, and eccentricity of bright points based on the entire dataset.

\subsubsection{Size distribution of filigree}

\begin{figure}
\centerline{
\includegraphics[width=1.0\textwidth]{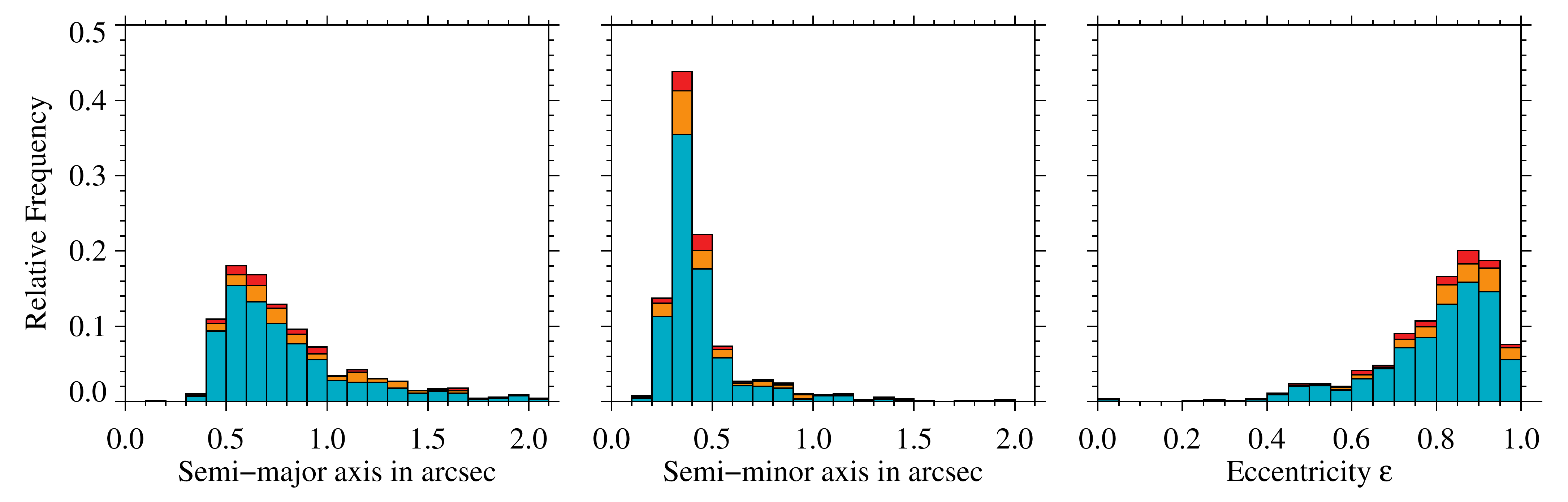}}
\caption{Histogram of semi-major axis (\textit{left}), semi-minor axis (\textit{middle}), and eccentricity (\textit{right}) of bright points. The histogram bins are divided into the three different classes: Type~I (\textit{cyan}), Type~II (\textit{orange}), and Type~III (\textit{red}). The bins are normalized to the total number of bright points ($n = 897$).}
\label{fig:stat}
\end{figure}

The filigree in this study are close to the solar polar regions. To study their properties such as length, width, or area, we have to correct the images geometrically by resampling the pixels to a regular grid with a pixel size of 71\,km, similar to the method described in \citet{Verma2011} for \textit{Hinode} images \citep{Kosugi2007}. This facilitates a more accurate estimation of the statistical parameters and makes it possible to compare our results with those obtained on other locations on the solar disk. To analyze the size distribution of filigree, we fit an ellipse to individual bright points and estimate their length and width based on the semi-major and -minor axis, respectively (right panel in Figure~\ref{fig:method}). The corresponding distributions are displayed in the left and middle panels of Figure~\ref{fig:stat} for each of the three types of filigree (see Section~\ref{s:method}).

The distribution of the semi-major axis of bright points is displayed in the left panel of Figure~\ref{fig:stat}. The peak of the distribution is between 0\arcsecdot5 and 0\arcsecdot6 for Type~I bright points, whereas Type~II and III bright points peak at about 0\arcsecdot6\,--\,\mbox{0\arcsecdot7}. The respective median values and related standard deviations for each type are $0\arcsecdot7\pm0\arcsecdot5$ (Type~I), $0\arcsecdot8\pm0\arcsecdot7$ (Type~II), and $0\arcsecdot7\pm1\arcsecdot0$ (Type~III). Higher values are reached only by few bright points with maximum values for of {5\arcsecdot1}, {5\arcsecdot4}, and {7\arcsecdot9} for Type~I, II, and III, respectively. These large values are likely related to (unresolved) cluster of bright points in each category. The minimal values are about 0\arcsecdot3 for Type~I and~II and about 0\arcsecdot2 for Type~III. The shape of the distributions indicates that most bright points are spatially resolved.

The histogram of the semi-minor axis shows a similar distribution (middle panel of Figure~\ref{fig:stat}), but the distribution is shifted towards lower values with a peak between \mbox{0\arcsecdot3}\,--\,0\arcsecdot4. The median and standard deviation for the three types are very similar with $0\arcsecdot38\pm0\arcsecdot21$ (Type~I), $0\arcsecdot39\pm0\arcsecdot28$ (Type~II), and $0\arcsecdot41\pm0\arcsecdot33$ (Type~III). Only a few bright points have a semi-minor axis above 1\arcsec\ and reach maximum values of 2\arcsecdot6 (Type~I) and about 1\arcsecdot9 for Type~II and III. The minimum values for the semi-minor axis are 0\arcsecdot18 for Type~I and III and 0\arcsecdot19 for Type~II. The minimum size of the bright points is limited by the selection procedure of bright points, where we excluded all structures with less than five pixels.

\begin{figure}
\centerline{
\includegraphics[width=1.0\textwidth]{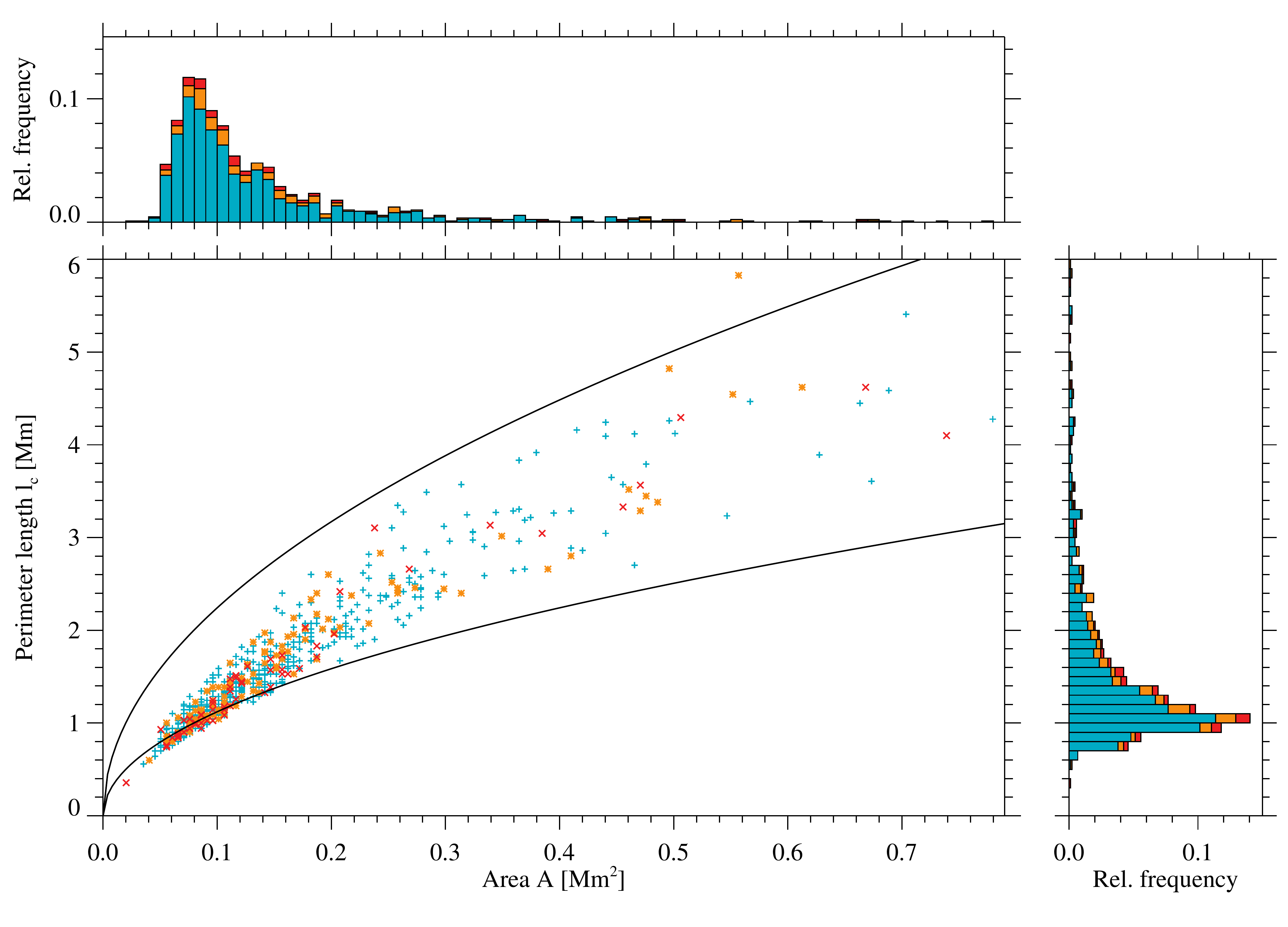}}
\caption{Area \textit{vs.} perimeter length and the respective histograms for bright points of Type~I (\textit{cyan} `$+$'), Type~II (\textit{orange} `$\ast$', and Type~III (\textit{red} `$\times$'). The envelope functions (\textit{black}) indicate $\alpha = 1$ and $\alpha = 2$ in Equation~(\ref{eq:pore}) for the lower and upper curve, respectively.}
\label{fig:stat_area}
\end{figure}

\subsubsection{Eccentricity of filigree}

The majority of bright points are best described by a circular shape but some bright points appear to have an elongated shape \citep{Muller2001}. In the following, we present a quantitative description of the bright points' shape approximated by the eccentricity $\epsilon$ of an ellipse following the approach of \citet{Verma2014}
\begin{equation}
\epsilon = \frac{e}{a} = \frac{\sqrt{a^2-b^2}}{a},
\end{equation}
where $e$ represents the linear eccentricity or distance of the focus points from the center, $a$ the semi-major axis and $b$ the semi-minor axis (right panel of Figure~\ref{fig:method}).

In the right panel of Figure~\ref{fig:stat}, a histogram of the eccentricity of bright points is shown. If the eccentricity is $\epsilon = 0$, the object has a circular shape and the semi-major and -minor axis are equal. Only three bright points have an $\epsilon = 0$ and 3.5\,\% of the bright points have an eccentricity of $\epsilon < 0.5$. The median eccentricity is 0.84, 0.86, and 0.83 with standard deviations of 0.14, 0.13 and 0.17 for Types~I, II, and III, respectively. This relates to an aspect ratio of about 2:1.

\subsubsection{Area distribution of filigree}

We displayed the area distribution of filigree, converted from pixels to megameters squared, in relation to the perimeter length in megameters after geometric correction of the images  (Figure~\ref{fig:stat_area}). In addition, we display the corresponding histograms differentiating the three categories of bright points discussed. The bright points have a median area of about 0.11\,Mm$^2$ with a standard deviation of 0.21\,Mm$^2$. The peak of the area distribution is about 0.08\,Mm$^2$ for the three types of bright points. The smallest area is 0.036\,Mm$^2$ (Type III) and the largest bright point or cluster thereof reaches an area of 3.5\,Mm$^2$ (Type~III). The perimeter length is increasing with increasing area. The median of the perimeter length is 1.24\,Mm with a standard deviation of 1.17\,Mm. The peak of the distribution of the perimeter length is between 1.0 and 1.2\,Mm for the three types. The minimal value for the perimeter length is reached for a Type~III bright point with 0.34\,Mm. The maximum perimeter length is 17\,Mm for a single very elongated bright point or cluster of bright points at a footpoint (Type~III). The majority of filigree have small areas and perimeter length and only few develop large areas with corresponding large perimeter length.

The following relation was defined for pores by \citet{Verma2014}
\begin{equation}
l_\mathrm{c} = 2\cdot \alpha \cdot \sqrt{\pi \cdot A}, \label{eq:pore}
\end{equation}
where $c$ is the perimeter length of the pore, $A$ is the area, and $\alpha$ is a constant, which describes how elliptical and jagged is a feature. The functions for $\alpha$ = 1 (assuming a circular object) and $\alpha$ = 2 (describing an elliptical or more rugged border) are displayed in Figure~\ref{fig:stat_area}. In most cases, Equation~(\ref{eq:pore}) describes the relation of the area to the perimeter length for bright points, but some outliers with very large perimeter length disturb the relation. This shows that some bright points have a very complex shape. For the lower approximation of $\alpha = 1$, the relation of area and perimeter length is not well approximated. The objects are too small, \textit{i.e.}, contain too few pixels for accurate ellipse fitting.

The three different groups of filigree are shown in Figure~\ref{fig:stat_area} with different symbols and colors. In general, the distribution of the bright points of Type~I is large, whereas the other two types cluster more in the lower end of the distribution with smaller areas and perimeter lengths. Nonetheless, there are individual Type~II and~III bright points which also possess large areas and moderate perimeter lengths.

\subsection{Filigree and their relation to magnetic field and UV intensity} \label{s:magfiligree}

\begin{figure}[t]
\centerline{
\includegraphics[width=0.95\textwidth]{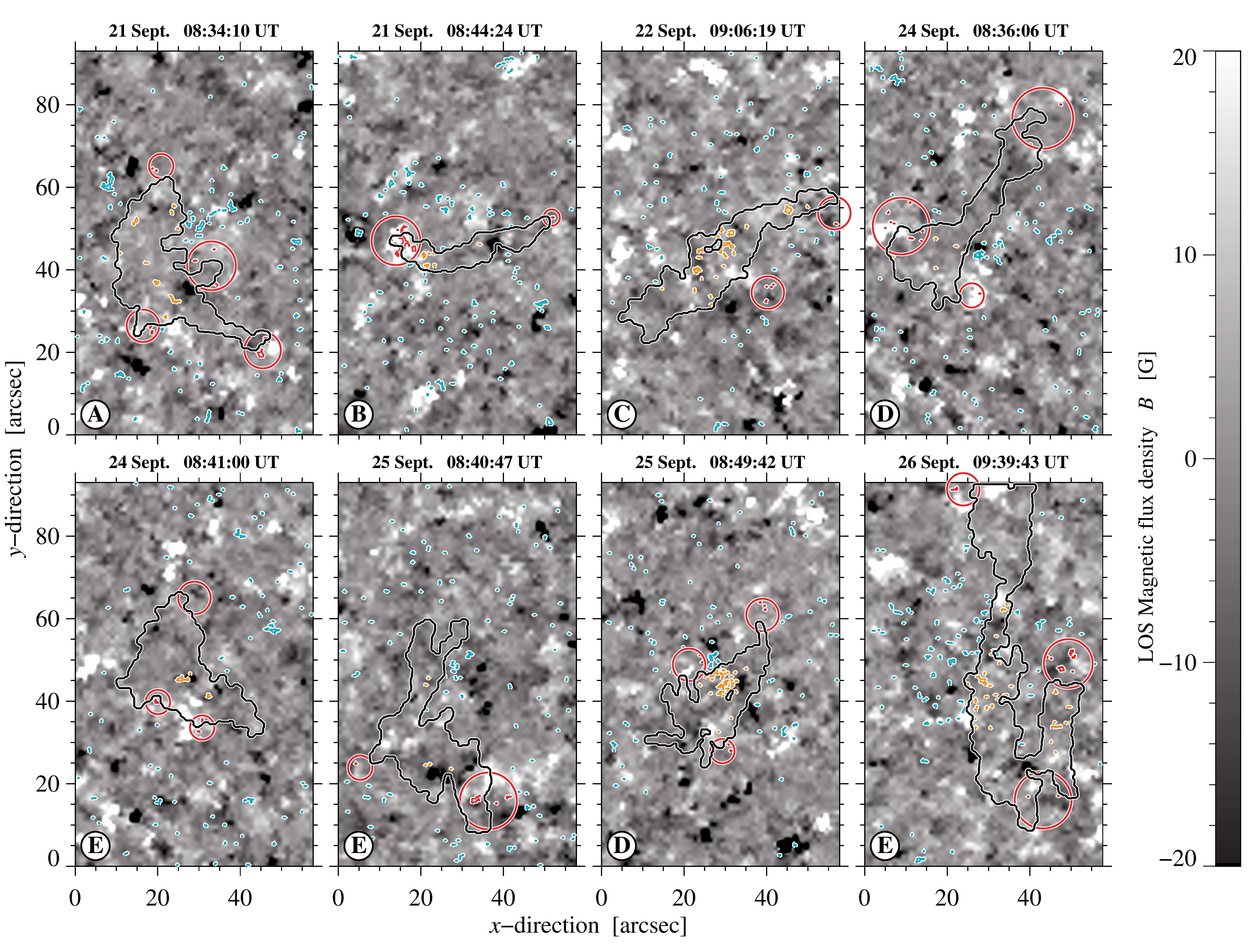}}
\caption{Deep HMI magnetograms, where the flux density is scaled between $\pm20$\,G with contours of filaments (\textit{black}) and filigree categorized as Type~I (\textit{cyan}), Type~II  (\textit{orange}), and Type~III (\textit{red}). In addition, the footpoints are marked with red circles. The label of the filaments is noted in a black circle in the lower left corner of each panel (see Table~\ref{tbl:overview}).}
\label{fig:mag_compare}
\end{figure}

The relationship between bright points and the magnetic field was discussed since the first description of filigree by \citet{Dunn1973}. We compare the location of the three types of filigree with the magnetic flux density (Figure~\ref{fig:mag_compare}). Most of the footpoint regions coincide with areas of enhanced magnetic flux density such as the left footpoint region of Filament~B in Figure~\ref{fig:mag_compare}. The footpoint is rooted in a positive-polarity patch, which is accompanied by a negative-polarity patch to the left (see also the magnified views in Figure~\ref{fig:bright}). In time-series of deep magnetograms, we observe that small-scale negative flux elements detach from the negative patch, which subsequently travel to the positive patch at the filament footpoint leading to flux cancellation. This occurs five times over a period of 10 hours. However, both the negative- and-positive polarity patch remain the dominant magnetic features in this region-of-interest over this 10-hour period, which indicates that the magnetic flux is intermittently replenished. We surmise that such a mechanism can provide the energy required to maintain and/or renew the cool plasma contained in the filament. We find also Type~I bright points (cyan contours in Figure~\ref{fig:mag_compare}) in the surroundings of the filaments, which coincide with concentrations of magnetic flux. Not all bright points are located in concentrations of magnetic flux but most are close to these areas. Filaments are located above the PIL and rooted in concentrations of magnetic field. Thus, we do not expect high photospheric magnetic flux density for Type~II bright points inside the filament contours. 

Filaments~D and~E (Figure~\ref{fig:mag_compare}) were observed on consecutive days and give some insight into the evolution of the photospheric magnetic field below the filaments over the course of two and three days, respectively. Filament~D is fully evolved on 24~September and decays on the following day. On the first observing day, the spine and three footpoints, rooted in strong concentrations of magnetic field, are clearly defined, whereas on the second day, the spine is not any longer elongated and distinct footpoints can no longer be recognized. Only the footpoint on the left is rooted in a region of high magnetic flux but no bright points are related to this footpoint region. The number of bright points at the location of the spine clearly increased on the second day. Filament~E significantly evolved during 24\,--\,26~September. On the first day, the filament has a more clumpy appearance, with three footpoints rooted in strong magnetic flux concentrations. On the second day, the footpoints are more difficult to define. In addition, a clear relation to high magnetic flux concentrations is missing. On the third day, the filament is more elongated, covering the entire FOV, and the single structure splits into two filaments, stretched out next to each other but connected by some threads (best seen in Figure~\ref{fig:overview_single}). The footpoints are not clearly visible in magnetograms. Not many bright points are found along the filament on the first two observing days but on the last observing day, their number increases. In full-disk observations (not shown), we see that Filament~E is decaying in the following days. The large number of bright points at the location of the filament can be taken as an indication that the filament starts to decay. However, to verify this hypothesis, more samples of filaments and bright points are needed on several observing days.

\begin{figure}[t]
\centerline{
\includegraphics[width=0.95\textwidth]{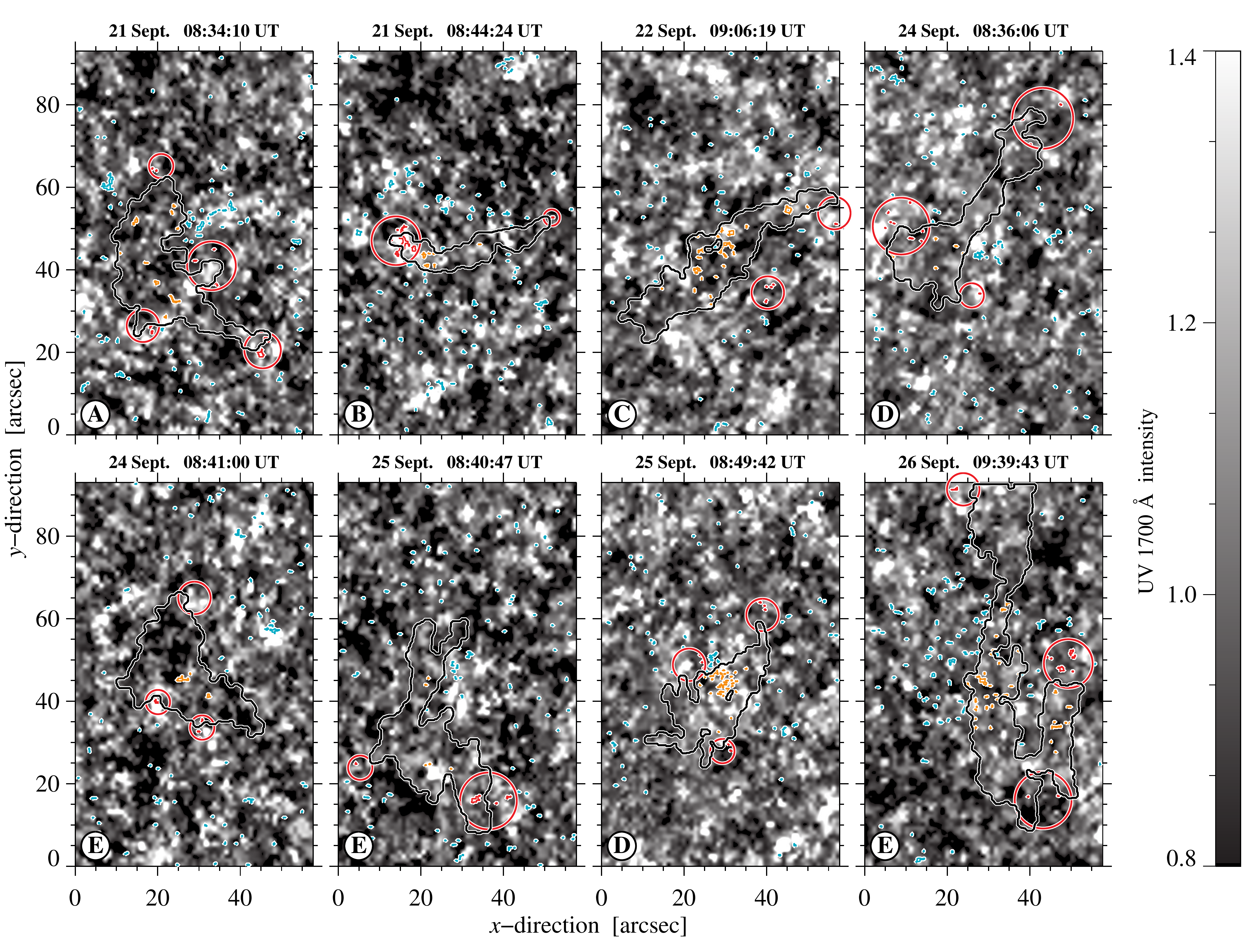}}
\caption{UV 1700\,\AA\ intensity of AIA normalized to the quiet-Sun intensity with the contours of the filaments (\textit{black}) and Type~I (\textit{cyan}), Type~II  (\textit{orange}), and Type~III (\textit{red}) filigree. In addition, the footpoints are marked with red circles. The label of the filaments is noted in a black circle in the lower left corner (see Table~\ref{tbl:overview}).}
\label{fig:1700_compare}
\end{figure}

In addition to magnetograms, we display the UV~1700\,\AA\ intensity (Figure~\ref{fig:1700_compare}), which we examine based on the position of bright points in H$\alpha$ images. The UV~1700\,\AA\ filtergrams are normalized to the median value of the quiet-Sun intensity $I_\mathrm{c}$ derived from the full-disk images, excluding bright and dark structures. Bright points are located close to intensity-enhanced regions in the UV~1700\,\AA\ filtergrams. The location of the bright points is related to the magnetic network in UV~1700\,\AA\ filtergrams. On 21~September, on the right side of Filament~A, bright points are clearly located at the border of a supergranular cell, including the bright points associated with the footpoints of the filament. Supergranular cell boundaries can also be determined for other filaments, \textit{e.g.}, on 24~September for Filament~D (lower left corner) and on 25~September for Filament~E (upper right corner). A similar behavior as for the UV~1700\,\AA\ is evident in \mbox{Ca\,\textsc{ii}\,K} filtergrams (Figure~\ref{fig:chrotel_compare}) and AIA UV~1600\,\AA\ filtergrams (not displayed).

\begin{figure}[t]
\centerline{
\includegraphics[width=1.0\textwidth]{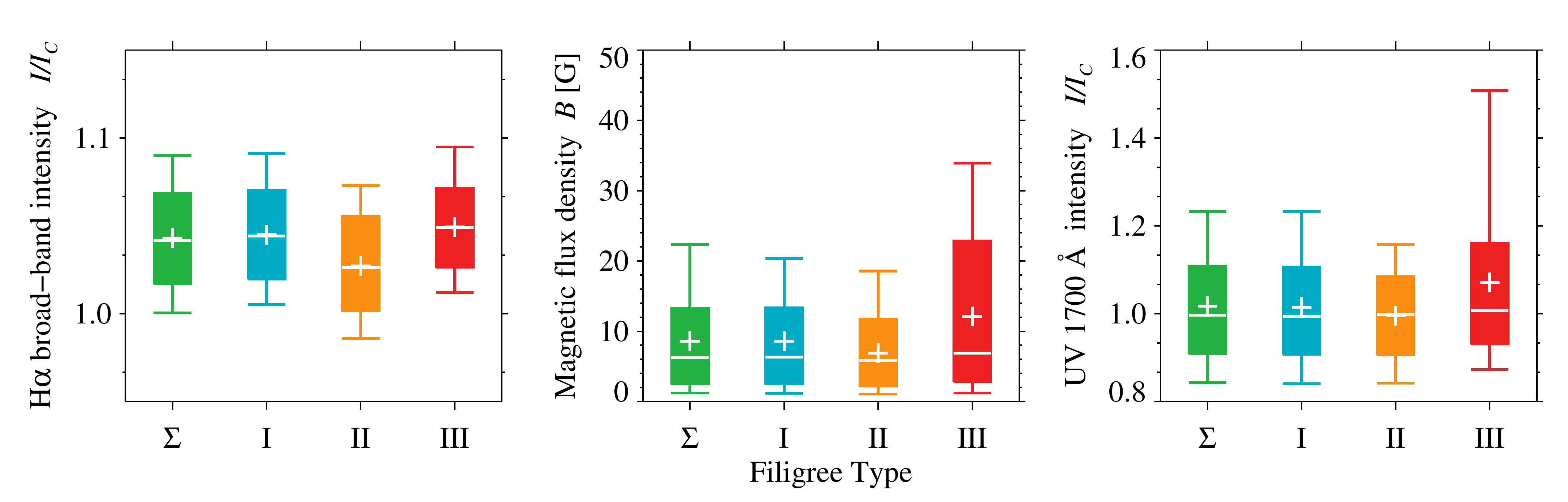}}
\caption{Box-and-whisker plot of the broad-band H$\alpha$ intensity (\textit{left}), magnetic flux density (\textit{middle}), and UV 1700\,\AA\ intensity (\textit{right}) for Type~I, II, and~III filigree (\textit{cyan, orange, and red}) and the aggregate (\textit{green}). The box is between percentiles P$_{16\%}$ and P$_{84\%}$, the whiskers between percentiles P$_{5\%}$ and P$_{95\%}$, the white line is the median value (P$_{50\%}$), and the `+' sign denotes the mean value.}
\label{fig:boxplot}
\end{figure}

We determined the intensity in the H$\alpha$ images and UV~1700\,\AA\ filtergrams as well as the magnetic flux density of the three types of bright points and display their statistics as a box-and-whisker plot (Figure~\ref{fig:boxplot}). Filigree are very bright in H$\alpha$ images but vary in intensity (left panel in Figure~\ref{fig:boxplot}). The intensity, normalized by the median value of the quiet-Sun intensity $I_\mathrm{c}$, is in the range 0.96\,--\,1.17\,$I/I_\mathrm{c}$ with a median and standard deviation of 1.04$\,I/I_\mathrm{c}$ and 0.03$\,I/I_\mathrm{c}$, respectively, for all bright points. Type~II bright points reach the lowest intensities of (1.026$\pm$0.027)$\,I/I_\mathrm{c}$ while Type~III bright points exhibit the highest intensities of (1.049$\pm$0.023)\,$I/I_\mathrm{c}$. For all three types, mean and median values are very similar, and the scatter is also insignificant. Most bright points are clearly brighter than unity, \textit{i.e.}, the median intensity of the quiet Sun that was used in the normalization.

 The magnetic flux density (middle panel in Figure~\ref{fig:boxplot}) of individual bright points was averaged across all pixels belonging to the bright point. The median values of the three types are very similar, \textit{i.e.}, around 6.0\,G, whereas the mean value varies more from 12.1\,G for Type~III, over 8.6\,G for Type~I, to 6.9\,G for Type~II. The scatter in the values is highest for Type~III bright points at the footpoints, which is also evident in the large differences for the percentiles. For Type~III, 84\,\% of the bright points have an average magnetic flux density of up to 22.9\,G, whereas for Type~I and Type~II, the values are about half of this with 13.4\,G and 11.8\,G. In addition, at the percentile P$_{95\%}$, 11\,\% of the bright points at the footpoints reach magnetic flux densities of up to 33.9\,G, whereas only 20.4\,G and 18.6\,G are reached for Type~I and~II, respectively. Furthermore, the standard deviation of 17\,G for Type~III shows a large scatter for the magnetic flux density at the footpoints. The standard deviation for Type~I is 10.0\,G and for Type~II it is only 5.0\,G. The maximum values are reached for Type~I and III with about 120\,G in both categories for individual bright points, whereas the maximum value of Type~II bright points is 26.0\,G. In summary, Type~III bright points, located at footpoints of polar crown and high-latitude filaments, are associated with higher concentrations of magnetic flux.
 
The box-and-whisker plot of the UV~1700\,\AA\ filtergrams (right panel in Figure~\ref{fig:boxplot},) shows that the median intensity is close to unity for the three types of bright points. The variation of the intensity is very low for Type~II bright points, where 95\,\% of the bright points have an intensity below 1.16\,$I/I_\mathrm{c}$, whereas for Type~III bright points, 95\,\% have an intensity below 1.51\,$I/I_\mathrm{c}$. In addition, the minimum intensity for this type is higher compared to the other types, with 5\,\% of the bright points having values below 0.87\,$I/I_\mathrm{c}$, whereas for the other types it is about 0.84\,$I/I_\mathrm{c}$. In summary, Type~III bright points at the footpoints are on average brighter than other bright points. This mirrors the relationship already found for the magnetic field.


\section{Discussion}\label{s:disc}


This rare sample of high-resolution polar crown and high-latitude filaments observed in H$\alpha$ images and filtergrams offered the opportunity to connect photospheric bright points with chromospheric filaments. We will discuss physical connections between filaments and bright points. Moreover, we address the question, if bright points in the weak magnetic fields of polar regions differ from their counterparts in the activity belt, which are well studied.

\subparagraph*{Morphology of filaments.} We first focus on the morphological description of the polar crown and high-latitude filaments recorded in high-resolution H$\alpha$ filtergrams. The high-resolution observations resolve many details such as single threads along the spine. In the surroundings, we identified mottles, some clustered around bright cores forming rosettes, and also some mini-filaments. The evolution of Filaments~D and~E was analyzed by snapshots in time over two and three days. Furthermore, detailed observations of the filaments showed that only one end was contained in the FOV for Filaments~A and~E, which led to the interpretation that these are fragments of filaments. Density structure and geometric considerations for high-latitude and polar crown filaments were already described in \citet{Schmieder2010} and \citet{Dudik2012}, who concluded that the viewing angle favors detecting cool plasma in a filament. Densities along the spine may be too low so that cool plasma escapes observations and leads to a fragmented appearance of filaments. Nonetheless, in full-disk H$\alpha$ and \mbox{He\,\textsc{i} filtergrams} more of the spine was visible and we could pinpoint the counterparts for the extreme ends nearby.

\subparagraph*{Irregular granulation.} The H$\alpha$ images are dominated by line-wing radiation, displaying photospheric granulation along with filigree. Some granules appeared brighter while other small areas appeared dark, which we attributed to strong absorption features and broad line profiles. In the intergranular lanes, we detected filigree as individual bright points or as clusters thereof. In the vicinity of bright points at lower latitudes, the granulation pattern often seems to be washed-out because of lower contrast and reduced visibility of the intergranular lanes, which is called ``abnormal granulation'' \citep{Dunn1973}. However, it is often related to strong magnetic flux concentrations in the vicinity of active regions \citep{Simon1974} close to pores \citep{Nesis2005} or sunspots \citep{Sobotka1994} with magnetic field strength of more than 100\,G \citep{Jin2009, Beck2017}. In this study, the LOS magnetic field does not reach 100\,G, except for some single pixels related to bright points, yet the typical pattern  of ``abnormal granulation'' is not encountered.

\subparagraph*{Morphology of filigree.} 

The bright points appear in different sizes and shapes and sometimes in clusters. The visual inspection of bright points led to the assumption that most bright points have a circular shape but many are significantly elongated \citep{Muller2001}. A cursory visual inspection of the sample in the present study supports this assumption but the quantitative analysis of the eccentricity shows a different result. Most of the bright points have an aspect ratio of about 2:1, and only three of the bright points have an eccentricity of nearly zero. Many bright points in H$\alpha$ form elongated chains, which shows a morphological similarity to bright points in \mbox{Ca\,\textsc{ii}}\,H \citep{Kuckein2009}. Nonetheless, not all bright points will be resolved. In an extensive statistical study, \citet{Yang2019} analyzed about 75\,000 bright points in \textit{Hinode} G-band images using a convolutional neural network. They differentiated between isolated points, elongated chains, and bright points in the form of a knee. Isolated points and elongated chains exhibited a mean eccentricity and standard deviation of 0.48$\pm$0.23 and 0.89$\pm$0.01, respectively. The latter is comparable to the median eccentricity in our sample of 0.83\,--\,0.86 for all three types.

The peak of the length distribution of the bright points resides in the range \mbox{0\arcsecdot5}\,--\,\mbox{0\arcsecdot7} with a median of the distribution of about 0\arcsecdot7. More elongated bright points have a length of up to 2\arcsecdot0 and only some clusters reach higher values. The median value of the width is about 0\arcsecdot4 for the three types of bright points. Earlier studies state a length of about 1\arcsecdot0\,--\,2\arcsecdot5 for elongated bright points \citep{Dunn1973, Tarbell1990} and diameters of 0\arcsecdot30\,--\,0\arcsecdot35 for circular bright points \citep{Muller2001, Kuckein2019}. Smaller diameters  between 0\arcsecdot13\,--\,0\arcsecdot16 are given by \citet{Berger2001}. In the study of \citet{Yang2019}, bright points have diameters in the range 0\arcsecdot14\,--\,0\arcsecdot43 with a mean diameter of 0\arcsecdot22 and a standard deviation of \mbox{0\arcsecdot03}, whereby the majority of bright points (88\,\%) have a circular shape. \citet{Leenaarts2006} concludes that bright points in the blue wing of H$\alpha - 0.08$\,\AA\ are predominantly photospheric and therefore have a similar morphology as G-band bright points. Since the broad-band filter covers the far wings of H$\alpha$ (Figure~\ref{fig:ha_trans}), we expect in these pseudo continuum images even more similarities to G-band bright points. However, our results indicate larger values for length and width of bright points observed at high-latitudes.

The area distribution was compared with the distribution of the perimeter length of bright points. A similar distribution was found for a statistical sample of pores from \textit{Hinode} observations by \citet{Verma2014}. They found that small pores have a tendency towards circular shapes and that larger pores tend to develop corrugated boundaries. A similar behavior is observed for bright points, where small bright points tend to be more circular whereas larger bright points are more elongated as seen in Figure~\ref{fig:stat_area}. This is reasonable because bright points are likely associated with single magnetic elements \citep[\textit{e.g.},][]{Muller2001}. Nonetheless, for smaller areas, close to the diffraction limit of the telescope, a reliable determination of the area and shape properties is impossible. On the other side of the scale, bright points with large perimeter length exhibit complicated structures. These are clusters of bright points, which were detected as one structure, \textit{e.g.}, near the left footpoint region of Filament~B (Figure~\ref{fig:bright}). One problem besides seeing and the telescope's diffraction limit is the tendency of pattern recognition algorithms to link neighboring features \citep{Verma2014}.

\citet{Crockett2010} reported area distributions of observed and simulated bright points with a peak at about 0.05\,Mm$^2$ and only a few exceeded 0.20\,Mm$^2$ in these observations obtained with the Dunn Solar Telescope (DST). The peak of the area distribution in our study is higher, \textit{i.e.}, in the range 0.07\,--\,0.09\,Mm$^2$. In addition, our study shows a large tail in the area distribution, up to about 0.50\,Mm$^2$. The larger area distribution can be attributed to the larger image scale of our observations (0.098\arcsec\,pixel$^{-1}$ at VTT \textit{vs.} 0.069\arcsec\,pixel$^{-1}$ at DST) or to the location on the solar disk (high latitude \textit{vs.} disk center) and consequently to projection effects. According to \citet{Yang2019}, circular bright points have a small area with a mean of about 0.02\,Mm$^2$, whereas more elongated bright points have mean areas of about 0.07\,Mm$^2$. In our sample, the peak in the area distribution was about 0.08\,Mm$^2$ but covers the range between 0.06\,Mm$^2$ and 0.11\,Mm$^2$ within one standard deviation, indicating a significant variation in size. Again, projection effects at high latitudes have to be considered.

The intensity of most bright points is clearly above the median intensity of the H$\alpha$ images, \textit{i.e.}, in the range 0.96\,--\,1.17\,$I/I_\mathrm{c}$, whereby Type~II bright points at the location of the filament tend to lower intensity values within this range with a median value of 1.03\,$I/I_\mathrm{c}$. Type~III bright points at the location of footpoints are in general brighter with a median intensity of 1.05\,$I/I_\mathrm{c}$. In the studies of \citet{SanchezAlmeida2004} and  \citet{Yang2019}, bright points reach intensities between 0.8\,--\,1.8 and 0.7\,--\,1.9, respectively, with the intensity normalized to the mean intensity of the quiet Sun. In the latter, the mean intensities of the identified bright points are 1.05 for circular bright points and reach higher values for more elongated bright point in the range 1.2\,--\,1.3.

\citet{Dunn1973} raised the question, if there are differences between the bright points in active regions, in the ``enhanced network'', and the quiet network. Most studies analyzed bright points at disk center \citep[\textit{e.g.},][]{Leenaarts2006} or active regions \citep[\textit{e.g.},][]{Kuckein2019}. Our study dealt with bright points at high latitudes in the vicinity of PCFs. The statistical evaluation, after careful geometrical correction to assure a correct interpretation of the morphology of the bright points, reveals that these bright points have very similar morphological properties as their counterparts at low latitudes but with larger length and width compared to lower latitudes.

\subparagraph*{Relations between filigree and filaments.} Surveying the evolution of the filigree inside the two Filaments~D and E over two and three days, we find that a large number of bright points are present before the decay of the filament. Since filigree are associated with individual magnetic elements, they may play an important role in the decay process by dispersing the remaining cool filament plasma. On the other hand, \citet{Li2016} reported bright points in the transition region at the base of filament threads, which potentially inject plasma into these threads. These bright points and the related upflows were quasi-periodic, resulting from small-scale oscillatory magnetic reconnection. In our study, we could not relate Type~II or III bright points to single threads. Therefore, there is no basis for claiming a relation with the mass supply via the bright points. Nonetheless, \citet{Engvold2004} and \citet{Lin2005} related the endpoints of single threads and barbs to  boundaries of supergranular cells  but they could not relate these endpoints to bright points. However, we find that bright points at footpoints of filaments are rooted in strong magnetic elements of the chromospheric network, which often correspond to the vertices of supergranular cell boundaries. The cool plasma is suspended above the weak fields inside supergranular cells, and the length of some predominate absorption structure in the observed filaments match the spatial scale of supergranulation.

\subparagraph*{Relations between filigree, magnetic field, and UV intensity.}
We related the location of filigree to the average magnetic flux density and UV intensity for the three types of bright points. This established that Type~III bright points at the footpoints are more likely to be found near higher magnetic flux concentrations and with enhanced UV intensities. About 95\,\% of the bright points in this category display magnetic flux densities of up to 34\,G. This indicates that bright points and filament footpoints are linked. In the first description of filigree, \citet{Dunn1973} stated that filigree are related to magnetic footpoints but this was neither followed up or extended to filaments of any type.

Some filigree appear in the vicinity of strong magnetic flux concentrations but they are not always co-spatial. The magnetograms will not show all quiet-Sun small-scale magnetic fields because of sensitivity limitations and the proximity close to the poles. Furthermore, we cannot rule out alignment and projection errors. \citet{Zhao2009} found that G-band and \mbox{Ca\,\textsc{ii}} bright points  were close to each other but did not necessarily intersect. \citet{Berger2001} examined the distance of G-band bright points to magnetic elements and found a mean distance of 0\arcsecdot24 but with most bright points intersecting with corresponding magnetic elements. Many studies stated that bright points are related to individual magnetic elements and that bright points in different wavelength bands are physically the same phenomenon, \textit{i.e.}, G-band, H$\alpha$, \mbox{Ca\,\textsc{ii}\,H}, \mbox{Ca\,\textsc{i}}~$\lambda$10839\,\AA, and the blue wing of \mbox{Fe\,\textsc{i}}~$\lambda$5250.5\AA\ and \mbox{Si\,\textsc{i}}~$\lambda$10827\,\AA\ \citep[\textit{e.g.},][]{Muller2001, Berger2001, Leenaarts2006, Zhao2009, Utz2014, Kuckein2019}.

The magnetic field values are relatively low for all filigree. However, the observations were carried out in quiet-Sun regions close to the poles. Other studies stating magnetic flux densities of bright points mainly focused on the activity belt around pores. Here, values of 200\,G to 1500\,G are encountered, especially in high-resolution observations of the magnetic field \citep[\textit{e.g.},][]{Tarbell1990, Muller2001, Beck2007, Utz2013, Kuckein2019}.

Our study supports the scenario that vertices of supergranular cells, \textit{i.e.}, the strongest and long-living magnetic flux concentrations in the quiet-Sun \citep{Giannattasio2018}, are often accompanied by bright points. Bright points are signatures of filament footpoints, which leave an imprint on the spatial scale of small high-latitude or polar-crown filaments or the building blocks of large-scale PCFs. In LOS magnetograms, we find intermittent flux cancellation at some of these sites, which provides energy to stabilize the mass-loaded filaments or to inject plasma into the filament \citep{Chae2000, Chae2003}. In addition, we observe an increasing number of bright points along the spine of a filament suggesting flux dispersal and a changing magnetic topology, which destabilizes the filament and initiates the decay of the filament. In general, morphological differences between bright points at low and high latitudes are small with the exception that high-latitude bright points are larger, which may be related to the weaker magnetic fields in polar regions.


\section{Conclusions and Outlook}\label{s:conc}


To our knowledge, this is the first high-resolution study examining bright points associated with polar crown and high-latitude filaments. They were extracted from high-resolution H$\alpha$ broad-band images and compared to their counterpart in H$\alpha$ narrow-band filtergrams and moderate-resolution magnetograms, UV~1700\,\AA\ filtergrams, and ChroTel full-disk filtergrams in H$\alpha$, \mbox{Ca\,\textsc{ii}\,K}, and \mbox{He\,\textsc{i}}. The main results can be summarized as follows: (i) The filaments and their surroundings could be resolved in great detail with high-resolution H$\alpha$ observations obtained with a new synchronized, dual-imager CMOS camera system. (ii) Bright points at high latitudes have a similar morphological appearance as bright points at low latitudes close to active regions but are slightly larger. (iii) We found that filigree at the footpoints are more likely located at stronger magnetic flux concentrations and that they exhibit enhanced UV intensities compared to other bright points in the surroundings. (iv) We observed a larger number of bright points at the location of the filament when the filament started to decay. This conclusion is based on the temporal evolution of two filaments followed over the course of two and three days. (v) Bright regions in UV~1700\,\AA\ and \mbox{Ca\,\textsc{ii}\,K} are clearly associated with bright points at the footpoints of filaments, in particular at the border of supergranular cells. Establishing the role of supergranular motions in forming polar crown filaments is a worthwhile scientific endeavor in itself but has to be deferred to a future investigation.

In the future, high-resolution fast cameras such as the M-lite 2M CMOS cameras, which were tested in this study, can be used for a detailed investigation of the temporal evolution of filigree and bright points. Time-series of polar crown filaments in H$\alpha$ broad- and narrow-band will resolve the interaction of small-scale magnetic features and filament fine structure, especially within the footpoint regions. This may hold clues for formation and decay of filaments and for mass supply and stability of long-lasting polar crown filaments. Additional high-resolution observations in other wavelength bands will provide a more comprehensive picture of bright points in different atmospheric layers. The synchronized M-lite 2M cameras are now part of the GFPI at the GREGOR solar telescope. In combination with the \textit{High-Resolution Fast Imager} \citep[HiFI:][]{Kuckein2017IAU} and \textit{GREGOR Infrared Spectrograph} \citep[GRIS:][]{Collados2012}, multi-wavelength observations, simultaneous magnetic field observations, and fast time-series become possible. New 4-meter class telescopes, such as the \textit{Daniel K.\ Inouye Solar Telescope} \citep[DKIST:][]{Tritschler2016DKIST} and the \textit{European Solar Telescope} \citep[EST:][]{Collados2010, Jurcak2019EST}, will soon provide data with a spatial resolution to fully resolve bright points facilitating a direct comparison with numerical simulations of small-scale magnetic features.


\begin{acks}
The Vacuum Tower Telescope (VTT) at the Spanish Observatorio del Teide of the Instituto de Astrof\'{\i}sica de Canarias is operated by the German consortium of the Leibniz-Institut f\"ur Sonnenphysik (KIS) in Freiburg, the Leibniz-Institut f\"ur Astrophysik Potsdam (AIP), and the Max-Planck-Institut f\"ur Sonnensystemforschung (MPS) in G\"ottingen. ChroTel is operated by the KIS in Freiburg, Germany, at the Spanish Observatorio del Teide on Tenerife (Spain). The ChroTel filtergraph was developed by the KIS in cooperation with the High Altitude Observatory (HAO) in Boulder, Colorado. Complementary H$\alpha$ full-disk data were provided by the Kanzelh\"ohe Solar Observatory, University of Graz, Austria. We acknowledge the support by grants DE~787/5-1 (CD, CK, MV) and VE~1112/1-1 (MV) of the German Research Foundation (DFG) and the support by the European Commission's Horizon 2020 Program (CD, CK, MV) under grant agreements 824064 (ESCAPE -- European Science Cluster of Astronomy \& Particle physics ESFRI research infrastructures) and 824135 (SOLARNET -- Integrating High Resolution Solar Physics). The authors thank Dr. J\"urgen Rendtel, Godehard Monecke, and Karin Gerber for their technical support during the observing campaign at VTT and Dr. Horst Balthasar for fruitful discussions. We thank Dr. Ioannis Kontogiannis for his helpful comments on the manuscript.
\medskip

\noindent\textbf{Disclosure of Potential Conflicts of Interest}$\quad$ The authors declare that they have no conflicts of interest. 
\end{acks}

%
%

\end{article} 
\end{document}